\documentclass[12pt,preprint]{aastex}

\usepackage{emulateapj5}           
\usepackage{natbib}

\slugcomment{Draft version from \today, accpeted for ApJL}

\shorttitle{The onset of massive star formation}
\shortauthors{Beuther et al.}

\begin{document}

\title{Caught in the act: The onset of massive star formation} 


\author{H.~Beuther$^1$, T.K.~Sridharan$^1$, M.~Saito$^2$}
\altaffiltext{1}{Harvard-Smithsonian Center for Astrophysics, 60 Garden Street, Cambridge, MA 02138, USA}
\altaffiltext{2}{National Astronomical Observatory of Japan, 2-21-1 Osawa, Mitaka, Tokyo, 181-8588, Japan}
\email{hbeuther@cfa.harvard.edu, tksridha@cfa.harvard.edu, Masao.Saito@nao.ac.jp}

\begin{abstract}
Combining mid-infrared data from the SPITZER Space Telescope with cold
gas and dust emission observations from the Plateau de Bure
Interferometer, we characterize the Infrared Dark Cloud IRDC\,18223-3
at high spatial resolution. The millimeter continuum data reveal
a massive $\sim$184\,M$_{\odot}$ gas core with a projected size of
$\sim$28000\,AU that has no associated protostellar mid-infrared
counterpart. However, the detection of 4.5\,$\mu$m emission at the
edge of the core indicates early outflow activity, which is supported
by broad CO and CS spectral line-wing emission. Moreover,
systematically increasing N$_2$H$^+$(1--0) line-width toward the mm
core center can be interpreted as additional evidence for early star
formation. Furthermore, the N$_2$H$^+$(1--0) line
emission reveals a less massive secondary core which could be in an
evolutionary stage prior to any star formation activity.
\end{abstract}


\section{Introduction} 

Massive star formation research has focused so far on evolutionary
stages either observable in the cm regime due to free-free emission
(Ultracompact H{\sc ii} regions, e.g.,
\citealt{kurtz2000,churchwell2002}), or on slightly younger stages
which emit in the (sub)mm bands and are also detectable at
mid-infrared (MIR) wavelengths due to warm dust emission (e.g.,
\citealt{molinari1996,sridha}). The initial phase prior to any cm
free-free and MIR warm dust emission was observationally
largely inaccessible. The most basic identification criterion for the
initial stages of massive star formation is that such regions have to
be bright in the (sub)mm regime due to cold dust and gas, and weak or
undetected in the MIR due to the absence or weakness of warm
dust emission. MIR surveys conducted with the Midcourse Space
Experiment (MSX) and the Infrared Space Observatory (ISO) permitted
identification of a large number of Infrared dark Clouds (IRDCs, e.g.,
\citealt{egan1998}). Studies of low-mass IRDCs have constrained the
initial conditions of low-mass star formation quite well (e.g.,
\citealt{andre2000,bacmann2000,alves2001,kirk2005}), but the high-mass
regime is just beginning to be explored (e.g.,
\citealt{garay2004,hill2005,klein2005,sridharan2005}).

In a 1.2\,mm continuum study of High-Mass Protostellar Objects (HMPOs)
associated with IRAS sources, we observed serendipitously within the
same fields additional mm-peaks not associated with any IRAS source
\citep{beuther2002a}. Correlation of these mm-peaks with the MSX MIR
data revealed a sample of mm-peaks that is not only weak in the MIR
but seen as absorption shadows against the galactic background
\citep{sridharan2005}. Mass estimates based on the 1.2\,mm continuum
emission show that they are massive gas cores (of the order a few
100\,M$_{\odot}$), making them potential High-Mass Starless Core
(HMSCs) candidates. Figure \ref{fig1} shows the HMSC candidate region
IRDC\,18223-3 which is a 1.2\,mm continuum peak at a distance of
$\sim$3.7\,kpc \citep{sridharan2005} in a dust and gas filament
approximately $3'$ south of IRAS\,18223$-$1243 (Peak \#3 in
\citealt{beuther2002a}). While the MIR absorption puts the source in
the sample of potential HMSCs, a NH$_3$ rotation temperature of
$\sim$33\,K indicates probable early star formation activity
\citep{sridharan2005}. The region was also identified recently as a
massive dense core by \citet{garay2004}.

\section{Observations}

We observed IRDC\,18223-3 with the Plateau de Bure Interferometer
(PdBI) during three nights in 2004/2005 at 93\,GHz in the C and D
configuration covering projected baselines between 15 and 230\,m. The
3\,mm receivers were tuned to the N$_2$H$^+$(1--0) line at
93.174\,GHz. The phase noise was lower than 30$^{\circ}$, and
atmospheric phase correction based on the 1.3\,mm total power was
applied. For continuum measurements we placed two 320\,MHz correlator
units in the band. The N$_2$H$^+$ lines were excluded in averaging the
two units to produce the final continuum image. Temporal fluctuations
of amplitude and phase were calibrated with frequent observations of
the quasars 1741$-$038 and 1908$-$201. The amplitude scale was derived
from measurements of MWC349. We estimate the final flux accuracy to be
correct to within $\sim 15\%$. The phase reference center is
R.A.[J2000] 18$^h$25$^m$08.3$^s$ and Dec.[J2000]
$-$12$^{\circ}$45$'$26.90$''$, and the velocity of rest $v_{\rm{lsr}}$
is 45\,km\,$s^{-1}$. The synthesized beam of the observations is
$5.8''\times 2.4''$ (P.A. $14^{\circ}$). The $3\sigma$ continuum rms
is 1.08\,mJy\,beam$^{-1}$. These mm observations are complemented with
the MIR SPITZER data from the GLIMPSE survey of the Galactic plane
using the IRAC camera centered at 3.6, 4.5, 5.8 and 8.0\,$\mu$m
\citep{werner2004,fazio2004,benjamin2003}. Observational details for
the CO(2--1) and CS(2--1) observations were given in
\citet{sridha,beuther2002a}.

\section{Results and Discussion}

Fig.~\ref{fig2} presents the MIR data toward IRDC\,18223-3 as a
3-color composite overlayed with contours of the 93\,GHz mm continuum
emission. We detect a compact dust and gas core spatially coincident
with the MIR dark lane, but we do not detect a (proto)stellar MIR
counterpart down to the sensitivity limit of this data ($3\sigma$:
0.05\,mJy@3.6\,$\mu$m, 0.05\,mJy@4.5\,$\mu$m, 0.13\,mJy@5.8\,$\mu$m,
0.15\,mJy@8.0\,$\mu$m). The projected size of the mm core in east-west
direction is $\sim$28000\,AU. Assuming the 93\,GHz continuum is due to
optically thin thermal dust emission, we estimate the mass and column
density of the core following \citet{beuther2002a}. Using the
temperature of 33\,K from NH$_3$ observations \citep{sridharan2005}
and a dust opacity index $\beta=2$, we calculate the mass within the
50\% contour level (integrated flux $S_{\rm{50\%}}\sim 6.7$\,mJy) to
be $\sim$95\,M$_{\odot}$, and the mass within the $3\sigma$ level of
1.08\,mJy ($S_{3\sigma}\sim 13.0$\,mJy) to be
$\sim$184\,M$_{\odot}$. The peak flux of $\sim$5.1\,mJy\,beam$^{-1}$
converts to a peak column density of $\sim 1.0\times
10^{24}$\,cm$^{-2}$, corresponding to a visual extinction of
$A_v=N_{\rm{H_2}}/0.94\times 10^{21}\sim 1000$
\citep{frerking1982}. The uncertainties for mass and column density
estimates from dust continuum emission are approximately within a
factor 5 (e.g., \citealt{beuther2002a}). Single-dish 1.2\,mm continuum
data \citep{beuther2002a} result with the same assumptions in a total
mass estimate of $\sim$245\,M$_{\odot}$, implying $\sim$25\% of
missing flux in the PdBI data due to missing short spacings. The data
clearly show that we are dealing with a massive gas core at an early
evolutionary stage.

While we do not detect a protostellar MIR source toward the mm-peak
IRDC\,18223-3, Fig.~\ref{fig2} shows weak emission in the SPITZER
4.5\,$\mu$m band (IRAC band 2 color-coded in green) toward the
south-east, the north-west and at the western edge of the mm
core. These features are just detected in the IRAC band 2. Since a
foreground source would also show up at 3.6\,$\mu$m, and an embedded
protostellar object would be red and thus also detectable at 5.8 and
8.0\,$\mu$m if PAH features are not too bright, these 4.5\,$\mu$m
features are unlikely to be of (proto)stellar nature. One cannot
entirely exclude that the 4.5\,$\mu$m emission is due to highly
reddened background sources (e.g., the MIR lower limit magnitudes
imply minimum visual extinctions $A_v>40$ and $A_v>120$ for the
south-eastern and the western 4.5\,$\mu$m features, respectively.),
but it is more likely that the 4.5\,$\mu$m features are due to line
emission within the IRAC band 2 bandpass (approximately 4 to
5\,$\mu$m). SPITZER outflow observations showed that molecular
outflows are particularly strong in the 4.5\,$\mu$m band because of
H$_2$ and CO line emission (e.g., \citealt{noriega2004}). Hence, we
suggest that the 4.5\,$\mu$m emission at the edge of the mm core may
be due to shock excitation as the outflow collides with the ambient
molecular medium. The two 4.5\,$\mu$m features to the south-east and
north-west are on opposite sides of the main mm-peak which is
indicative of typical bipolar outflows \citep{richer2000}. The
additional third feature to the west is suggestive of multiple
outflows as often observed in massive star-forming regions (e.g.,
\citealt{beuther2003a}).

\begin{figure}
\begin{center}
\includegraphics[angle=-90,width=7cm]{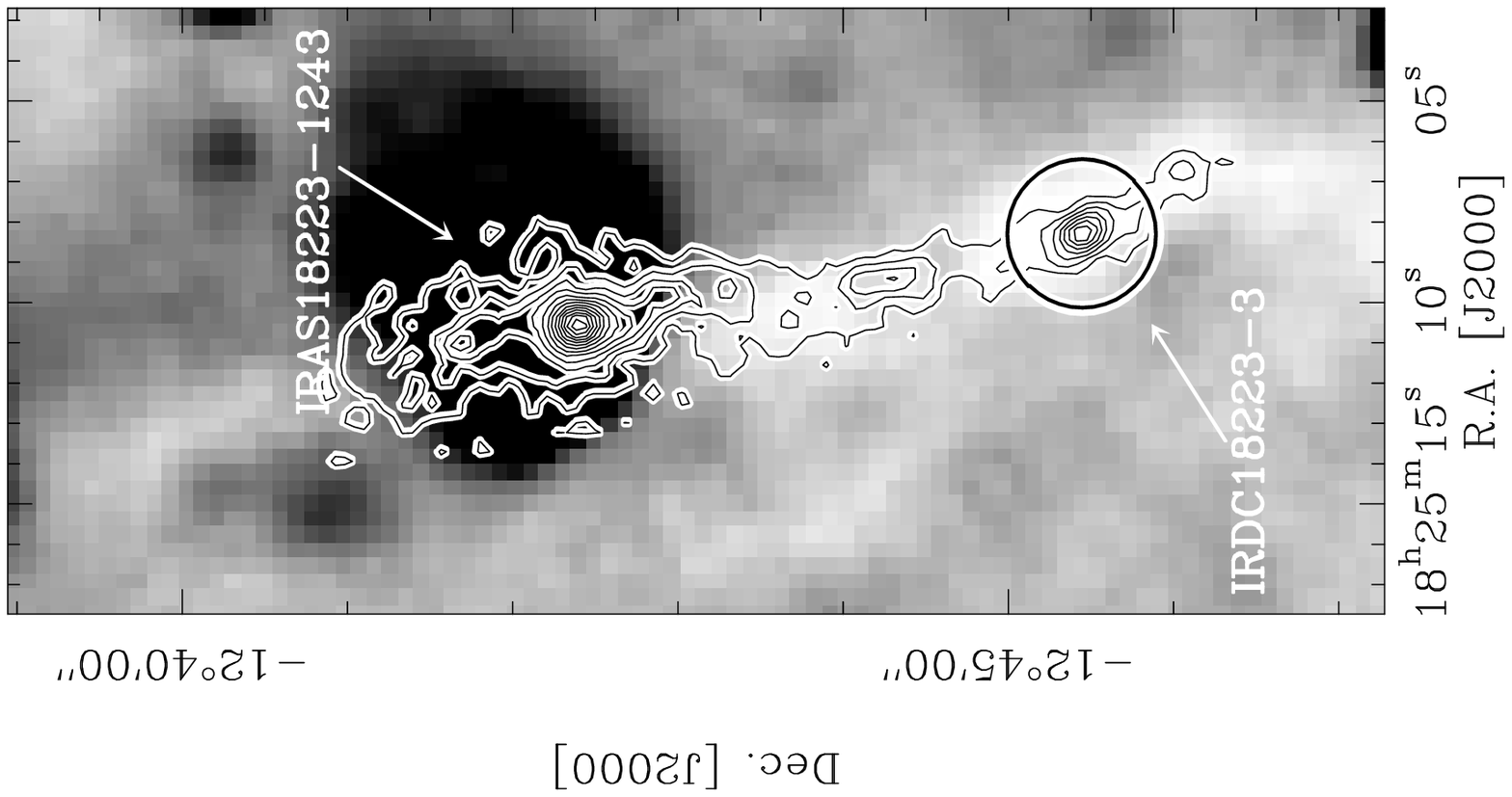}
\end{center}
\caption{Contour overlay of the 1.2\,mm single-dish continuum map
\cite{beuther2002a} on the 8\,$\mu$m MSX image (in grey-scale). The
contour levels are from 38\,mJy in 38\,mJy steps. The northern source
is the High-Mass Protostellar Object (HMPO) IRAS\,18223-1243, and the
southern source is the High-Mass Starless Core candidate
IRDC\,18223-3. The black circle outlines the primary beam of the
Plateau de Bure Interferometer observations.}
\label{fig1}
\end{figure}

\begin{figure}
\begin{center}
\includegraphics[width=8cm]{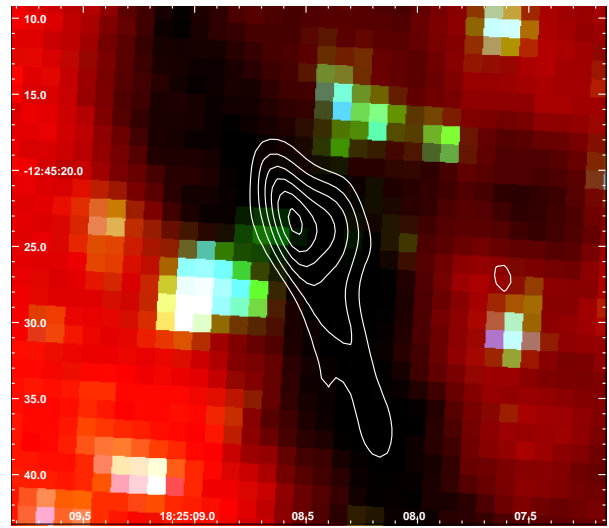}
\end{center}
\caption{Three color image of the SPITZER IRAC observations at 3.6
(blue), 4.5 (green) and 8.0\,$\mu$m (red). The contours show the PdBI
93\,GHz continuum map from 1.08\,mJy ($3\sigma$) in 0.72\,mJy steps
(2$\sigma$). The axis are in R.A. [J2000] and Dec. [J2000].}
\label{fig2}
\end{figure}

\begin{figure}
\begin{center}
\includegraphics[angle=-90,width=8cm]{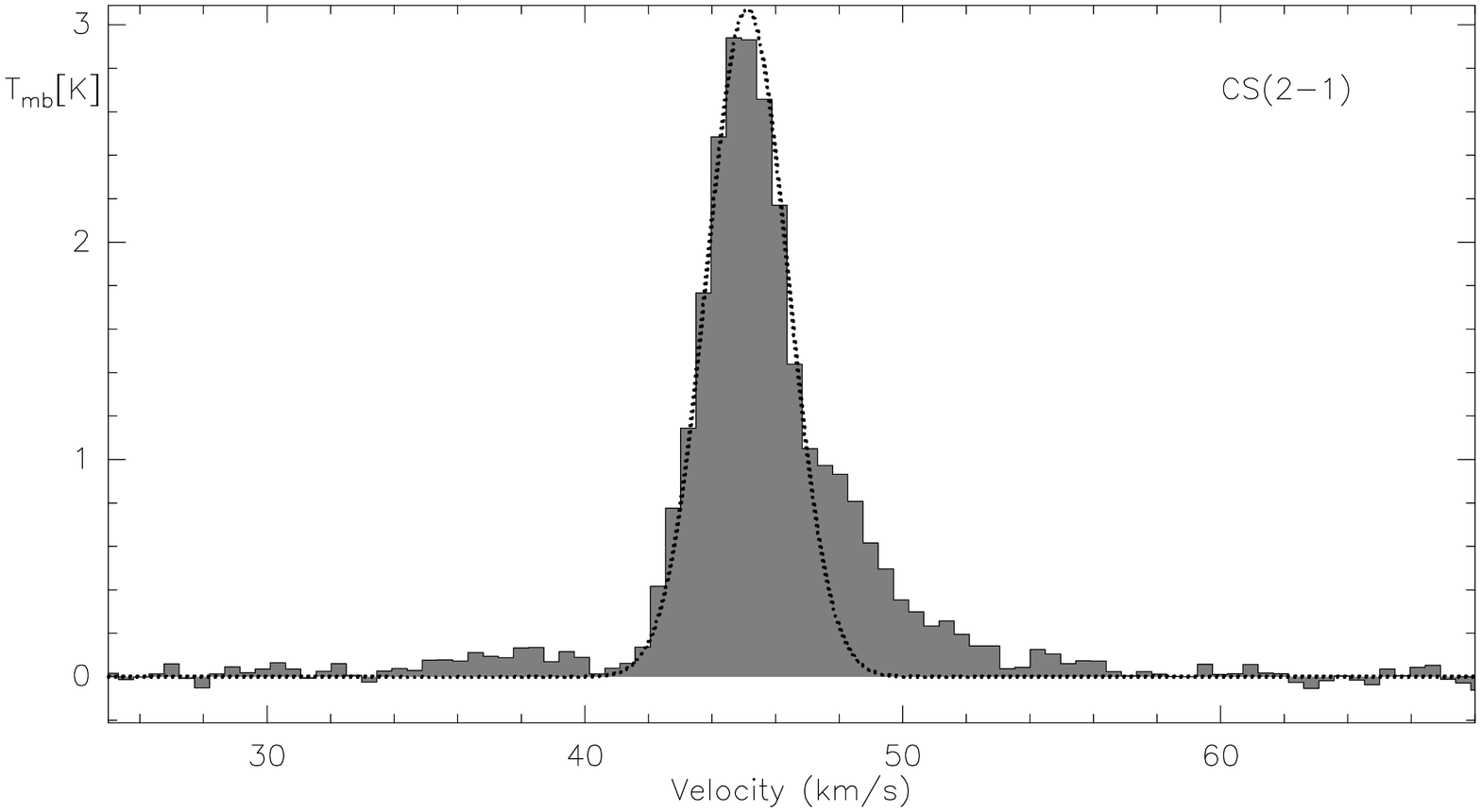}
\includegraphics[angle=-90,width=8cm]{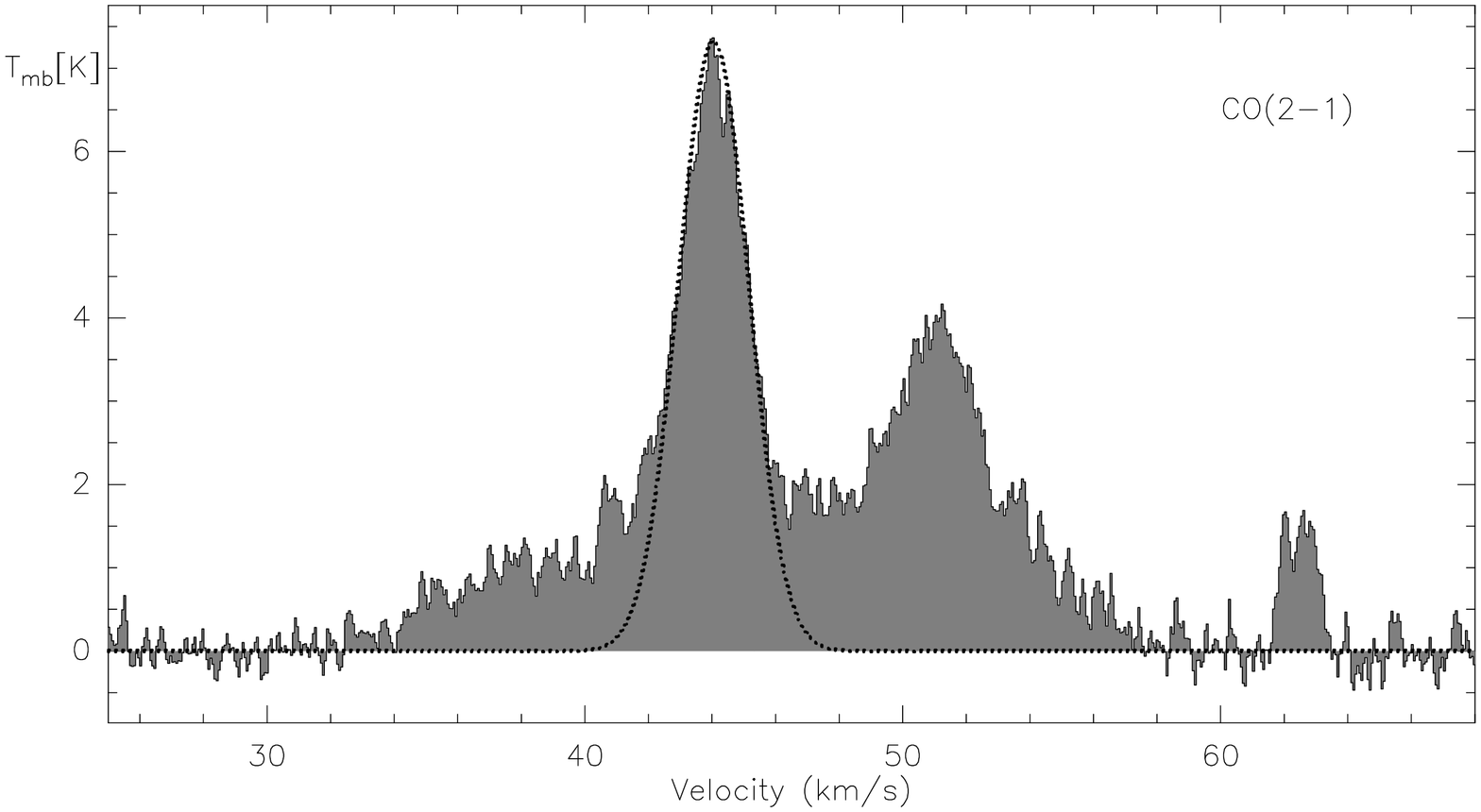}\\
\includegraphics[angle=-90,width=8cm]{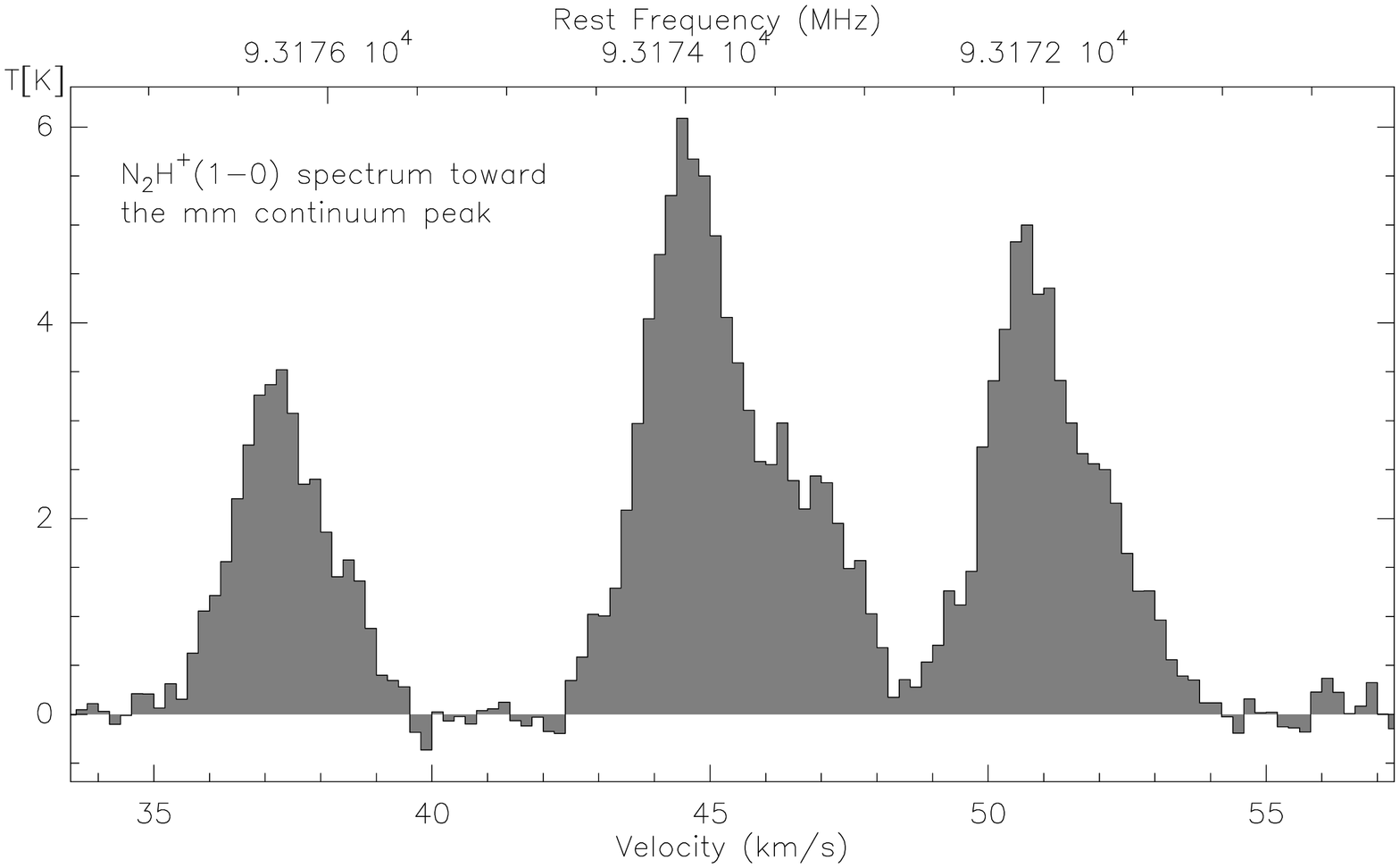}
\includegraphics[angle=-90,width=8cm]{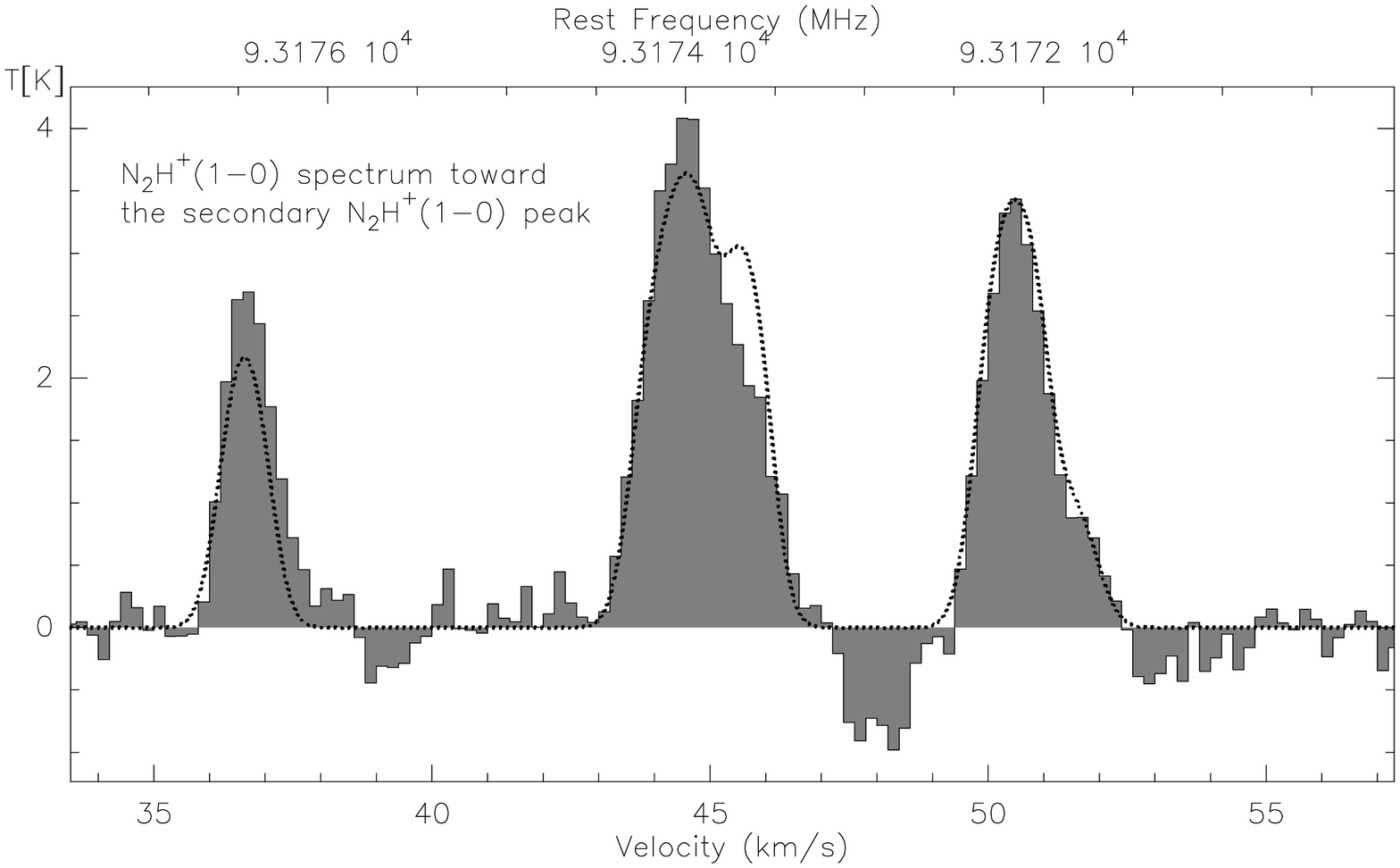}
\end{center}
\caption{{\bf Top:} Single-dish CS(2--1) and CO(2--1) spectra toward
IRDC\,18223-3 observed with the IRAM 30\,m and CSO, respectively. The
solid lines show the spectra and the dotted lines Gaussian fits to the
central line component at $v_{\rm{lsr}}\sim 45$\,km\,s$^{-1}$.
Obviously, both spectra show strong excess line wing emission
indicative of a high-velocity molecular outflow. {\bf Bottom:}
N$_2$H$^+$(1--0) spectra in IRDC\,18223-3 toward the 93\,GHz continuum
peak (left) and the secondary N$_2$H$^+$(1--0) peak $\sim$7$''$ to the
east (right) observed with the PdBI. Note that the presented velocity
range is smaller than for the CO/CS spectra. The N$_2$H$^+$ spectra
cover all seven hyperfine components, the rest-frequency is set to
the hyperfine group frequency of 93173.770\,MHz. The dotted line in
the right spectrum shows the fit to the hyperfine
structure. The negative features in the right spectrum are not real
absorption features but due to the negative side-lobes caused by the
missing short spacing data in the corresponding velocity channels
toward this position.}
\label{fig2b}
\end{figure}

Support for this outflow interpretation is provided by single-dish
CO(2--1) and CS(2--1) observations from the Caltech Submillimeter
Observatory (CSO) and the IRAM 30\,m telescope
\citep{sridha,beuther2002a}. Both spectra presented in
Fig.~\ref{fig2b} show line-wing emission in excess of a Gaussian line
component at the $v_{\rm{lsr}}\sim 45$\,km\,s$^{-1}$. The Full Width
at Zero Intensity of both lines is $\sim$24\,km\,s$^{-1}$. The
CO(2--1) spectrum shows two additional line components at $\sim$51
and $\sim$62\,km\,s$^{-1}$. Comparing these features with CO(2--1)
spectra toward other positions of the large-scale dust filament
(Fig.~\ref{fig1}), the 51\,km\,s$^{-1}$ component is also present
toward the rest of the filament whereas the 62\,km\,s$^{-1}$ component
is observed only toward IRDC\,18223-3. This difference indicates that
the 52\,km\,s$^{-1}$ component is probably part of a larger-scale
foreground or background cloud, whereas the 62\,km\,s$^{-1}$ component
is likely due to the molecular outflow. Following \cite{beuther2002b},
we roughly estimate the mass of high-velocity gas within the CSO
primary beam ($31''$) to be $\sim$3.4\,M$_{\odot}$. Compared to other
typical high-mass outflows (e.g., \citealt{beuther2002b}), this is a
relatively low value. However, since we are presumably dealing with a
source at the onset of massive star formation, there has not been much
time yet to eject and entrain molecular gas, and thus comparably low
outflow masses are expected at this evolutionary stage.

\begin{figure}
\begin{center}
\includegraphics[angle=-90,width=15.8cm]{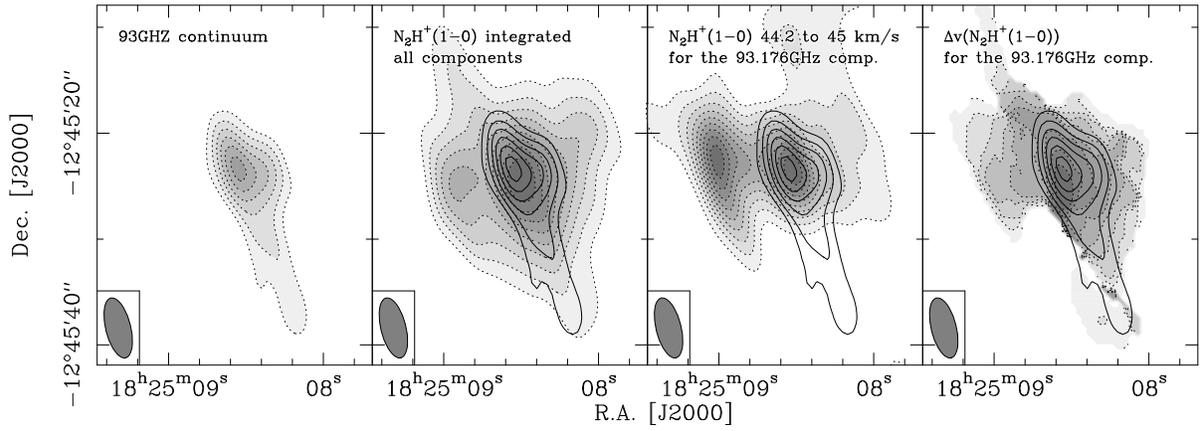}
\end{center}
\caption{93\,GHz continuum and N$_2$H$^+$(1--0) emission and
line-width maps toward IRDC\,18223-3. The grey-scale with thin 
dotted contours shows the
93\,GHz continuum and N$_2$H$^+$(1--0) maps as labeled within each
panel.  The solid contours in panels 2 to 4 again show the 93\,GHz
continuum emission. The contouring of the continuum starts at
1.08\,mJy ($3\sigma$) and proceeds in 0.72\,mJy steps (2$\sigma$).
The contour levels of the N$_2$H$^+$(1--0) maps range always from 10
to 90\,\% of the peak values (23.5\,Jy\,beam$^{-1}$, 1.2\,Jy\,beam$^{-1}$ and
2.0\,km\,s$^{-1}$, respectively). The synthesized beam is shown at
the bottom-left of each panel.}
\label{fig3}
\end{figure}

In addition, we observed the N$_2$H$^+$(1--0) line with seven
hyperfine components around 93.174\,GHz (Fig.\ref{fig2b}), which is
known to be strong and optically thin in low-mass starless cores
\citep{tafalla2004}. Fig.~\ref{fig3} presents the N$_2$H$^+$(1--0)
emission integrated over all hyperfine components, and over a smaller
velocity interval centered on the isolated N$_2$H$^+$(1--0) component
at 93176.27\,MHz. Both N$_2$H$^+$ maps show an emission peak toward
the main IRDC\,18223-3 mm continuum peak. Moreover, the N$_2$H$^+$
data exhibit a secondary peak $\sim$7$''$ to the east of this main
peak, spatially associated with the MIR absorption north-east of the
4.5\,$\mu$m emission.  Furthermore, we find that the N$_2$H$^+$
line-width $\Delta v$ increases from the outer edges of the N$_2$H$^+$
emission in the direction toward the main IRDC\,18223-3 mm-peak
(Fig.~\ref{fig3}).  Although the main IRDC\,18223-3 peak likely
exhibits multiple velocity components (see below), the general trend
of increasing $\Delta v$ appears real. This is indicative of increased
internal motion within the main IRDC\,18223-3 core~-- either turbulent
or ordered motion like infall, rotation or outflow. We interprete this
as additional evidence for the onset of star formation activity. The
line-width is still narrow toward the IRDC\,18223-3 secondary
N$_2$H$^+$ peak with $\Delta v \sim 1$\,km\,s$^{-1}$, indicating less
internal motion and hence likely an earlier evolutionary stage.

Fitting the N$_2$H$^+$ hyperfine structure permits determination of
the optical depth and N$_2$H$^+$ column density (e.g.,
\citealt{caselli2002a}). Unfortunately, we cannot derive a good fit
toward the main IRDC\,18223-3 mm-peak because the N$_2$H$^+$ spectrum
shows excess emission 1-2\,km\,s$^{-1}$ offset from the peak velocity
(Fig.~\ref{fig2b}) indicative of multiple velocity components. The
spectral fitting difficulties are likely due to this complex velocity
structure. In contrast, we can fit the spectrum toward the secondary
IRDC\,18223-3 N$_2$H$^+$ peak reasonably well (Fig.~\ref{fig2b}). The
resulting N$_2$H$^+$ column density of $3.1\times 10^{13}$\,cm$^{-2}$
translates into an H$_2$ column density of $N_{\rm{H_2}}\sim 1.0\times
10^{23}$\,cm$^{-2}$, assuming a N$_2$H$^+$/H$_2$ ratio of $3\times
10^{-10}$ \citep{caselli2002b}. Since the 3\,mm continuum
$3\sigma$\,rms of 1.08\,mJy corresponds to an H$_2$ column density
sensitivity between $2.2\times 10^{23}$ and $5.2\times
10^{23}$\,cm$^{-2}$ (assuming 33 and 15\,K, respectively), the
continuum non-detection of this N$_2$H$^+$ secondary peak is a
plausible observational result. Assuming a lower temperature of 15\,K
(typical for HMSC candidates, \citealt{sridharan2005}), the $3\sigma$
mm continuum sensitivity implies that a core as massive as
37\,M$_{\odot}$ can be hidden toward this secondary N$_2$H$^+$ peak
without being detected in our mm continuum data.

We estimate the virial masses of the IRDC\,18223-3 main and secondary
N$_2$H$^+$ peaks for the observed line-widths $\Delta v$ (2.0 and
1.0\,km\,s$^{-1}$, Fig.~\ref{fig3}) and sizes $R$ ($\sim 14000$\,AU
for both peaks) following \citet{maclaren1988}. Assuming different
density distributions ($\rho\propto 1/r$ \& $\rho\propto 1/r^2$), the
derived virial masses are 51 \& 34 and 13 \& 8\,M$_{\odot}$ for the
two N$_2$H$^+$ peaks, respectively. This estimate for the main
N$_2$H$^+$ peak is more than a factor 3 lower than what we derived
from the 93\,GHz mm continuum emission. Although the error budget is
high for the various mass estimates (about a factor 5,
\citealt{beuther2002a}), the smaller virial mass compared to the gas
mass derived from the mm continuum is consistent with this core
collapsing to form a star. The virial mass of the secondary N$_2$H$^+$
peak is consistent with the previously derived upper mass limits from
the mm continuum. Hence, the secondary core may still be in a virially
bound state, potentially prior to active star formation.

In summary, the observational features of (i) a massive IRDC, (ii) a
molecular outflow as suggested by the 4.5\,$\mu$m, CO and CS emission,
(iii) the N$_2$H$^+$ line-width and virial analysis, and (iv) the high
NH$_3$ temperatures indicate the presence of an extremely young
massive protostellar object at the center of the mm continuum core
IRDC\,18223-3, which remains otherwise undetected in the MIR
due to its extreme youth and the high gas column densities.  Although
this line of evidence is still circumstantial and has to be further
investigated, the MIR non-detection and the comparably low
outflow mass supports an early evolutionary stage prior to the typically
studied HMPOs. The narrow linewidth and associated virial mass of the
lower-mass secondary N$_2$H$^+$ peak suggests that it may be in an
even younger pre-protostellar stage prior to any active star
formation.

\acknowledgments{IRAM is supported by INSU/CNRS (France), MPG
(Germany), and IGN (Spain). SPITZER is operated by the JPL, Caltech
under NASA contract 1407. We wish to thank J.\,Hora, C.\,de Vries,
J.\,Joergensen, T.\,Bourke, T.\,Megeath and T.\,Huard for help with
the SPITZER and N$_2$H$^+$ data. We also thank the referee for
detailed comments improving the paper. H.B. acknowledges financial
support by the Emmy-Noether-Programm of the Deutsche
Forschungsgemeinschaft (DFG, grant BE2578).  }


\clearpage

\clearpage

\clearpage

\clearpage

\end{document}